\documentclass{elsart}
\usepackage{graphicx}
\usepackage{xspace}

\usepackage{amssymb}

\begin{document}

\begin{flushright}
Fermilab-Pub-05-045-AD
\end{flushright}
\vskip .5 cm

\begin{frontmatter}



\title{Beam Test of a Segmented Foil SEM Grid}


\author[ut]{S.E. Kopp\corauthref{myemail}}
\corauth[myemail]{Corresponding author e-mail kopp@mail.hep.utexas.edu} 
\author[ut]{D.Indurthy}
\author[ut]{Z.Pavlovich}
\author[ut]{M.Proga}
\author[ut]{R.Zwaska}
\author[fnal]{S.Childress}
\author[fnal]{R.Ford}
\author[fnal]{C.Kendziora}
\author[fnal]{T.Kobilarcik}
\author[fnal]{C.Moore}
\author[fnal]{G.Tassotto}
 
\address[ut]{Department of Physics, University of Texas, Austin, Texas 78712 USA}
\address[fnal]{Fermi National Accelerator Laboratory, Batavia, Illinois 60510 USA}

\begin{abstract}
A prototype Secondary-electron Emission Monitor (SEM) was installed in the 8~GeV 
proton transport line for the MiniBooNE experiment at Fermilab.  The SEM is a segmented
grid made with 5~$\mu$m Ti foils, intended for use in the 120~GeV NuMI beam 
at Fermilab.  Similar to previous workers, we found that the full collection of the secondary electron signal requires a bias voltage to draw the ejected electrons cleanly off the foils, and this effect is more pronounced at larger beam intensity.  The beam centroid and width resolutions of the SEM were measured at beam widths of 3, 7, and 8~mm, and compared
to calculations.  Extrapolating the data from this beam test, we expect a centroid and width resolutions 
of $\delta x_{beam}=20~\mu$m and $\delta \sigma_{beam}=25~\mu$m, respectively, in the NuMI beam which has 1~mm spot size.

\end{abstract}

\end{frontmatter}

\section{Introduction}
\label{sec:intro}

Beam profiles may be measured via the process of secondary electron emission\cite{bruining}.    A secondary electron monitor (SEM) consists of a metal screen of low work function from which low ($<$100 eV) energy electrons are ejected.  While the probability for secondary electron emission is low ($\sim$0.01/beam particle), these devices can produce signals of 10-100nC when $4\times10^{13}$ beam particles per spill pass through the device, permitting their use as beam monitors\cite{tautfest}.  Furthermore, the process of secondary electron emission is a surface phenomenon\cite{surface}, so that electron emitting foils or wires of very thin (1-10~$\mu$m) dimensions may be used without penalty to the signal size\cite{anne}.  Often a positive voltage on a nearby foil ("clearing field") is used to draw the secondary electrons cleanly away from the signal screen.  A schematic SEM is shown in Figure 1.

Secondary electron emission monitors have replaced ionization chambers as beam monitors for over 40 years\cite{tautfest}.  An ionization chamber monitors beam intensity by measuring the ionized charge in a gas volume collected on a chamber electrode.  Such a device places a large amount of material ($\sim 10^{-2}-10^{-3}~\lambda_{int}$) in the beam which results in emittance blowup and beam loss, both of which are unacceptable in high intensity beams.  A further limitation of ionization chambers is that space charge buildup limits them to measurements of beams with intenstities of  $<10^{16}$~particles/cm$^2$/sec\cite{ion}, nearly 5 orders of magnitude below requirements of present extracted beamlines.   SEM's, in contrast, are extremely linear in response \cite{tautfest,tainuty}, and the prototype SEM discussed in this note is $7\times10^{-6}$ interaction lengths thick.

\begin{figure}[t]
  \begin{centering}
  \includegraphics[width=3in]{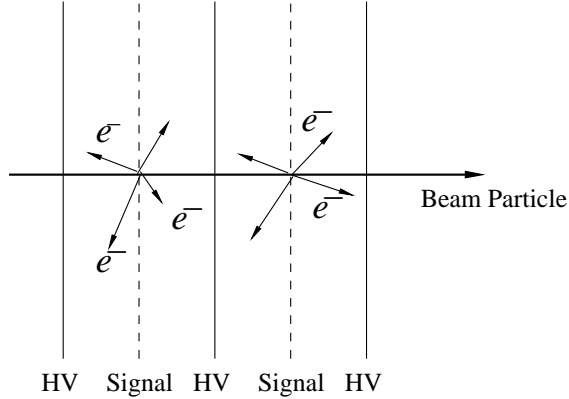}
  \caption{Schematic of a segmented secondary emission monitor (SEM):  electrons ejected by the signal planes are drawn away by bias planes at a positive voltage. 
  \label{fig:semconcept}}
  \end{centering}
\end{figure}

For the NuMI beam \cite{numi}, we desire a segmented SEM which measures the beam intensity, the beam centroid position, and the beam's lateral profile.  The beam spot is anticipated to be $\sim$1mm.  The required SEM segmentation is of order 1mm.  The two SEM's near the NuMI target require segmentation of 0.5mm in order to specify the beam position and angle onto the target at the 50~$\mu$rad level.  The segmented SEM will measure profile out to 22~mm in the horizontal and vertical.  A single, large foil will cover the remaining aperture out to 50~mm radius in order to measure any potential beam halo.  Additional thin foil SEM's are envisaged for the 8~GeV transport line for the MiniBooNE experiment \cite{miniboone} and for the transfer line between the 8~GeV Booster and 120~GeV Main Injector at FNAL.

A prototype SEM was tested in the 8~GeV beam transport line for the MiniBooNE experiment in May 2003. While the MiniBooNE beam parameters differ from those anticipated for NuMI (see Table~\ref{tab:MiniBooNEvsNuMI}), this test permitted early verification of the foil SEM design.  Some differences, listed in Table~\ref{tab:MiniBooNEvsNuMI}, exist between the foils designed for the prototype and the final SEM chambers installed in the NuMI line. Further details of the prototype design are given in Section~\ref{sec:proto}, while the final SEM design description can be found in Ref~\cite{SEMFinal}.  

During the beam tests, the SEM was used to measure beam position and size at one location in the MiniBooNE line.  Because it was the only profile monitor in that portion of the transport line, no independent measurement existed to corroborate the prototype's beam size measurements.  A pair of nearby capacitative Beam Position Monitors (BPM's) was able to corroborate the SEM's beam centroid measurement.  The SEM's expected beam centroid resolution and beam width resolution are related, however, because both depend upon several aspects of the SEM design, such as readout noise and the position accuracy of the segmented SEM grid assembly.  In this note, we analyze the SEM's centroid and width resolution during the test in the MiniBooNE line.  These measurements are compared to calculations of expected centroid resolution performance.  Following the validation of the calculations using the test beam data, we extrapolate the expected beam size resolution achievable from the SEM's to the case for the narrow NuMI beam.

\begin{table}[!h]
\begin{center}
\begin{tabular}{|l|c|c|}
\hline
BEAM                                   & MiniBooNE & NuMI      \\
\hline
Proton energy (GeV)                    & 8         & 120        \\
Intensity ($\times10^{12}$ ppp)        & 5         & 40        \\
Spill Rate (Hz)                      & 5         & 0.5       \\
Spill Duration ($\mu$s)               & 1.56      & 8.67      \\
Horizontal beam size (mm)            & 6-8   & 1.0   \\
Vertical beam size (mm)              & 3   & 1.0 \\
\hline
SEM                               & Prototype & Final SEM       \\
\hline
Strip width (mm)                       &   0.75    &  0.15       \\
Strip pitch (mm)                       &     1     &  1             \\
Foil thickness ($\mu$m)               &     5     &  5    \\
\hline
\end{tabular}
\caption{Comparison of MiniBooNE and NuMI beam lines and of the prototype and final design SEM's. Characteristics of the prototype that was tested in MiniBooNE beamline along with the characteristics of the final design SEMs that are used in NuMI beamline are listed. 
\label{tab:MiniBooNEvsNuMI}}
\end{center}
\end{table}

\section{SEM Prototype}
\label{sec:proto}

\begin{figure}[t]
  \begin{centering}
  \includegraphics[width=.85\textwidth]{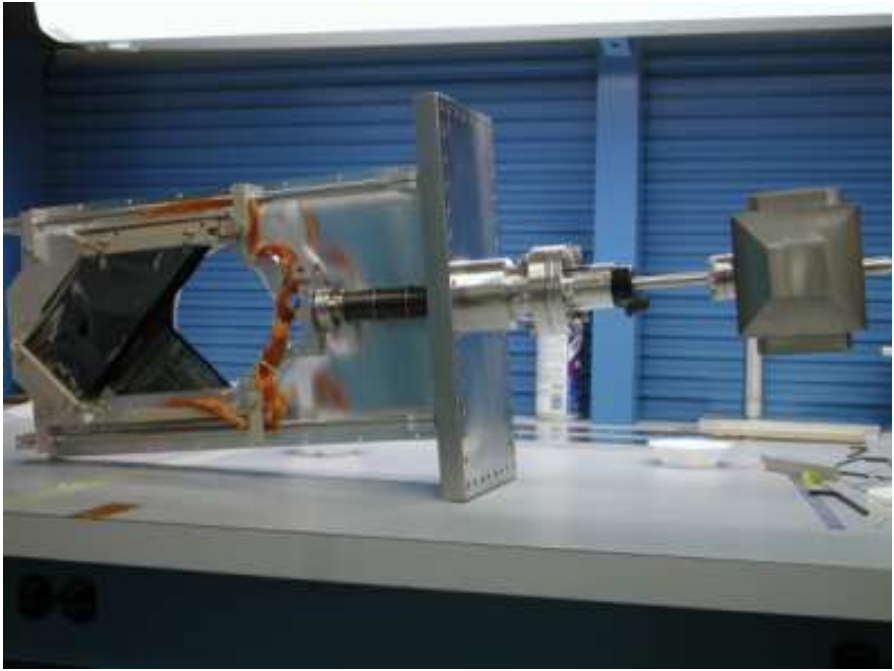}
  \caption{Photograph of the prototype segmented Foil SEM.  At right is the signal connection feedthrough box.  At center is the lid of the vacuum chamber for the SEM, and at left is the paddle with the foils.  The foil paddle moves in and out of the beam on rails, and the vacuum during the motion is maintained by a bellows feedthrough mounted to the vacuum chamber lid. The assembly shown is approximately 75~cm long by 30~cm tall.
  \label{fig:semphoto}}
  \end{centering}
\end{figure}

\begin{figure}[b]
  \begin{centering}
   \includegraphics[width=.75\textwidth]{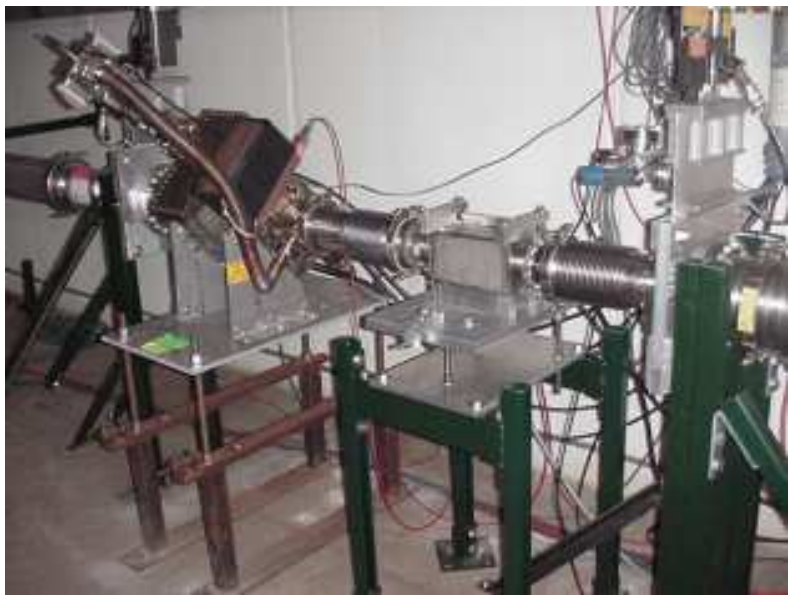}
  \caption{Photograph of the prototype segmented Foil SEM installed in the 8~GeV beamline.   
  \label{fig:semphoto2}}
  \end{centering}
\end{figure}

Borrowing from a design in use at CERN \cite{ferioli}, the foil SEM built for this beam test had five planes of 5~$\mu$m thick Titanium foils, as in Figure~\ref{fig:semconcept}.  The first, third, and fifth planes were solid foils biased to as much as 100~Volts.  Planes 2 and 4 were segmented at 1~mm pitch (0.75~mm wide strips with 0.25~mm gaps between adjacent strips).  The strips were mounted on ceramic combs
with rectangular grooves which mechancally held the strips and aligned them onto the grid.  Each strip had an accordion-like spring to tension it and compensate for beam heating.  The foil strips for the prototype were quite long, ranging from 15-25~cm in length.  Each strip was read out separately into a charge-integrating circuit\cite{swic} gated around the beam spill.

\begin{figure}[t]
  \begin{centering}
  \includegraphics[width=4in]{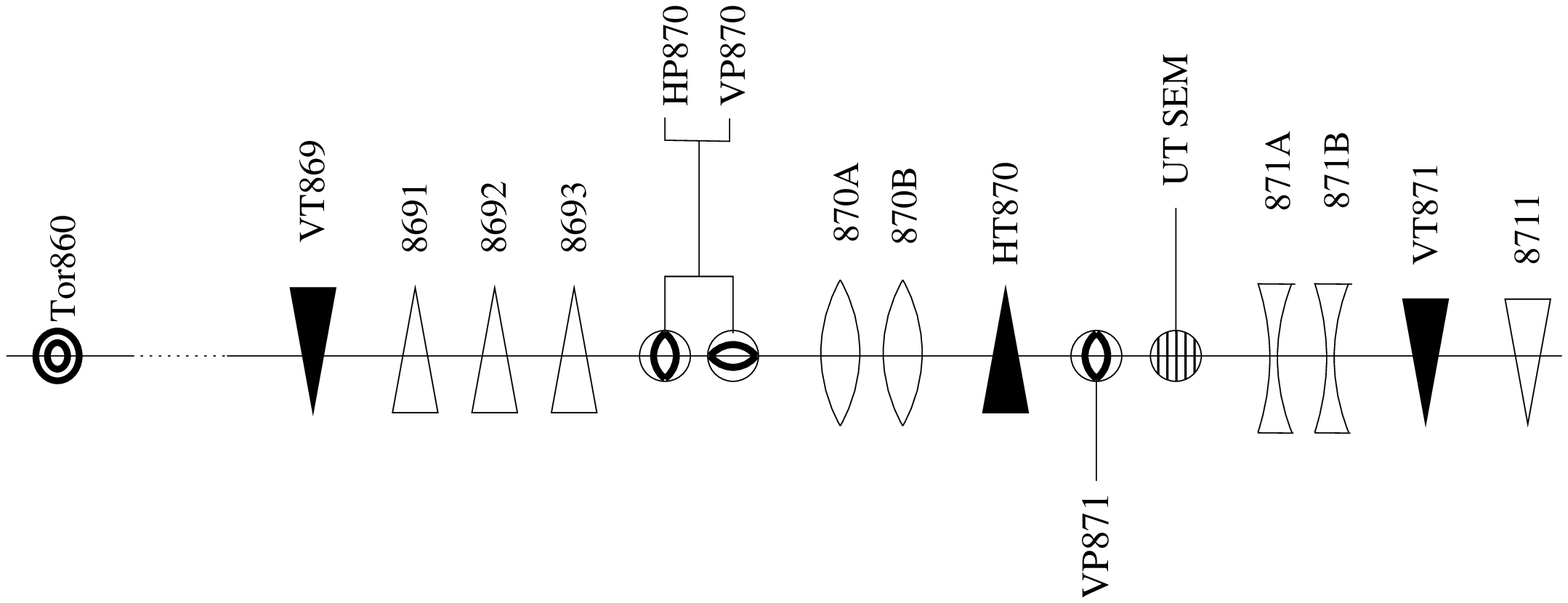}
  \caption{Beamline segment around SEM prototype. Tor~=~beam current toroid, HP and VP~=~horizontal and vertical beam position monitor (BPM), HT and VT~=~horizontal and vertical trim magnets, UT SEM~=~University of Texas foil SEM.
  \label{fig:Beamline}}
  \end{centering}
\end{figure}

A new aspect of the present SEM design is that the foils were mounted on a frame which does not traverse the beam as the SEM is inserted into or retracted from the beam.  The segmented foils are mounted at $\pm45^\circ$ to provide both horizontal and vertical beam profiles, and these are mounted on a hexagonal frame which encloses the beam at all times, leaving a clear space when it is desired to retract the foils away from the beam, as shown in Figure~\ref{fig:semphoto}.  The signals from the foils are routed through a vacuum bellows feedthrough via kapton-insulated cables to a signal feedthrough box shown at the right of the photo.  While the final NuMI SEM chambers have stepper motor-driven actuators \cite{SEMFinal} to move the foils into or out of the beam, the prototype required manual manipulation of the bellows feedthrough to insert the foil paddle.  A photograph of the SEM, mounted in its rectangular vacuum chamber and installed in the 8~GeV transport line, is shown in Figure~\ref{fig:semphoto2}.

The use of Titanium as the active medium was motivated by the loss of Secondary Electron Emission (SEE) signal from other materials after prolonged exposure in the beam \cite{isabelle,garwin,agoritsas,ferioli2}.  For beam nominally on center through a long run, such signal loss results in degraded beam centroid magnitude and also results in artificially enhanced beam tails, since the beam tails irradiate the SEM to a lesser extent.  Titanium suffers less from the loss of SEE signal, even for relatively simple cleaning and handling procedures for the foils \cite{ferioli2}.

\section{Data analysis}
\label{sec:analysis}
Figure~\ref{fig:Beamline} shows a schematic of the 8~GeV MiniBooNE transport line elements near the foil SEM location. To analyse the SEM data we wanted to correlate it with data from nearby BPM's. However, there is no simple linear relation between all the BPM and SEM data. Changing magnet currents, beam position and intensity causes both the slope and the offset in the relation between SEM and BPM to change. The situation is worse for horizontal than for vertical plane since there are magnets in between the SEM prototype and horizontal BPM. 

The  data, consisting of $\sim$80,000 spills, was split into periods during which the currents in nearby dipole, quadrupole, and trim magnets, as well as the beam intensity, were relatively constant. The beam current transformer Tor860 was used for intensity monitoring. Only the spills with sufficiently high intensity ($\sim10^{12}$~ppp) were analysed.

\begin{figure}[b]
  \begin{center}
  \includegraphics[width=0.47\textwidth]{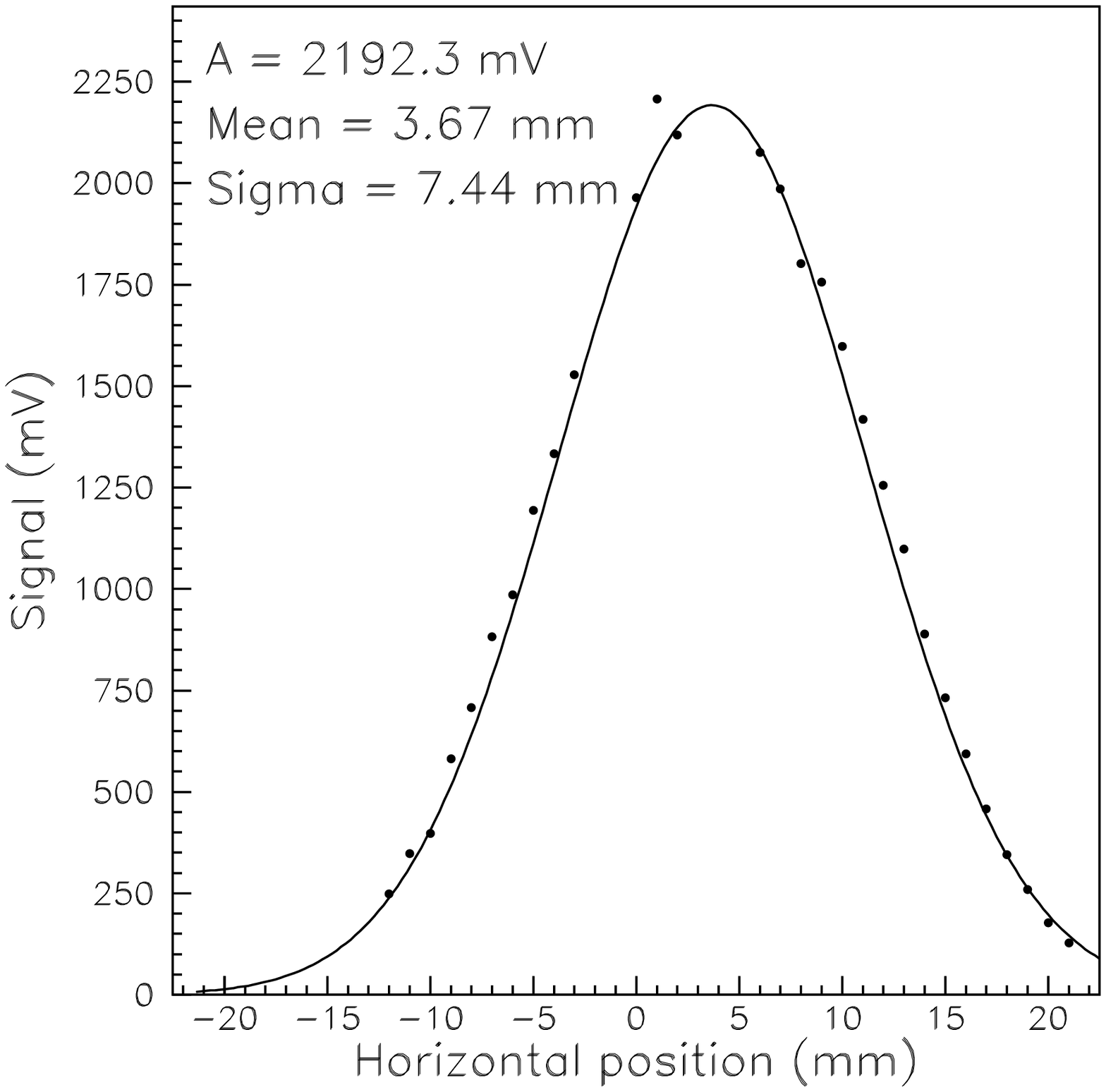}
  \includegraphics[width=0.47\textwidth]{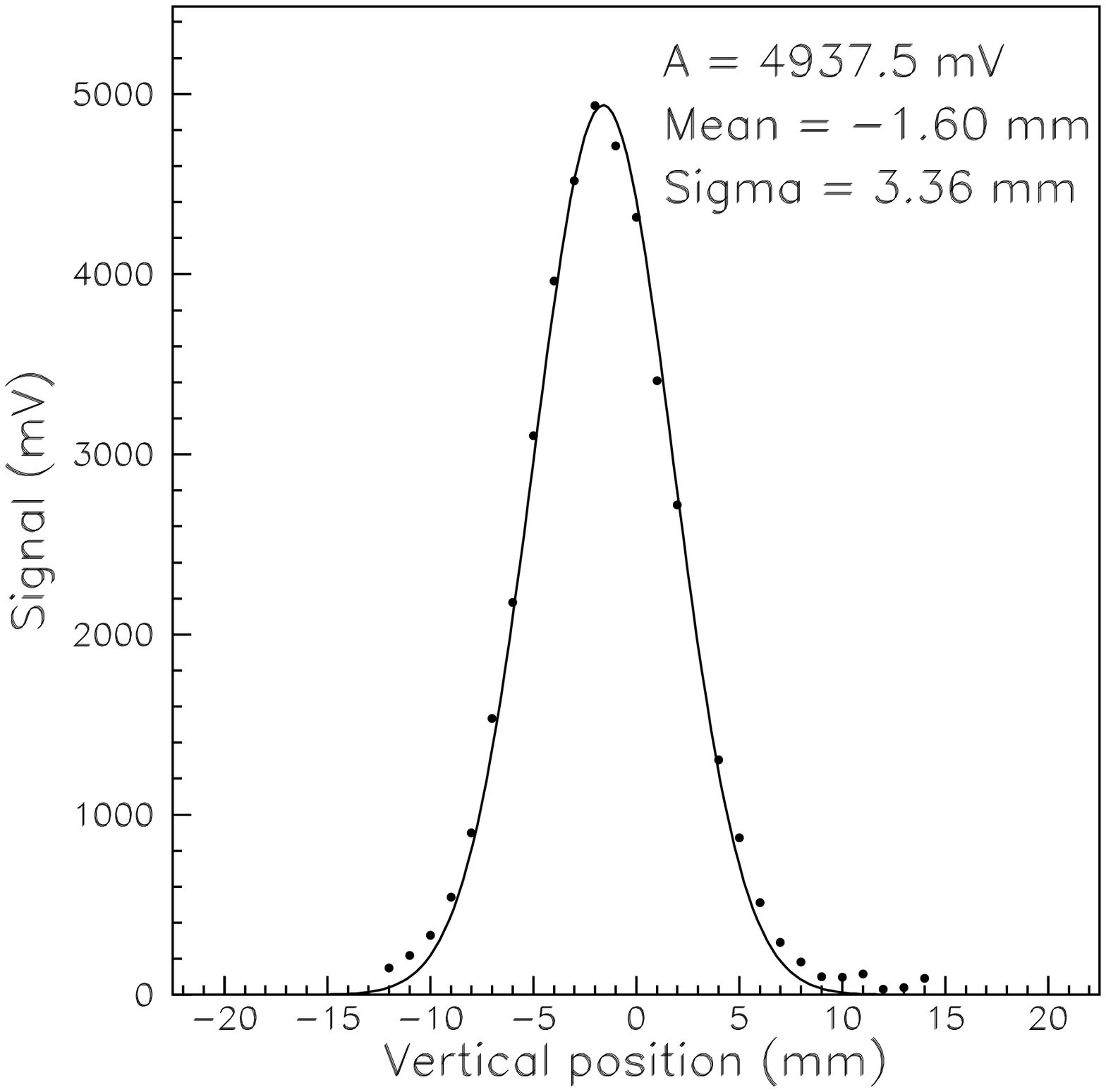}
  \caption{Horizontal and verical beam profiles during one spill at $4.5\times10^{12}$~ppp.  The fitted beam 
centroid, width and amplitude are noted in the plots.  
  \label{fig:beamprofile}}
  \end{center}
\end{figure}

Figure~\ref{fig:beamprofile} shows one example of the horizontal and vertical beam profiles as seen by the SEM in one spill with $4.5\times10^{12}$~protons per pulse (ppp). The vertical axis on these plots is the pulse height from a given strip\cite{caveat}.  The beam was broader in the horizontal direction and narrower in the vertical direction. Both horizontal and vertical beam widths roughly agree with expectations based on the $\beta$ functions for the transport line. 

\begin{figure}[t]
  \begin{center}
  \includegraphics[width=.8\textwidth]{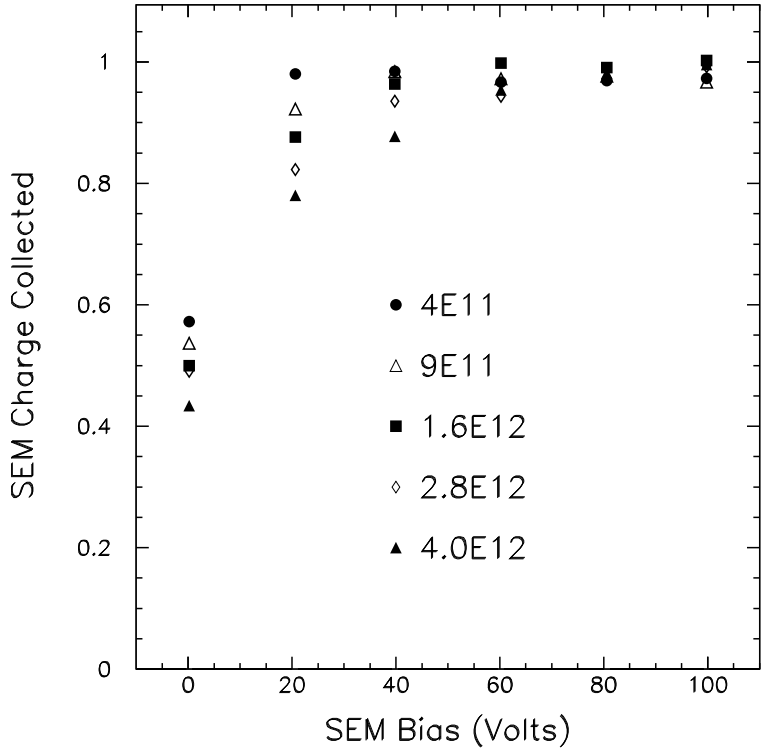}
  \caption{(Normalized) total charge collected from the SEM summed over all 44 strips, plotted as a function of the applied bias voltage.  The data were taken at several beam intensities, ranging from 4$\times10^{11}$~protons/pulse to 4$\times10^{12}$~protons/pulse.}
\label{fig:biasvolts}
  \end{center}
\end{figure}

\section{Bias Voltage Study}
The voltage applied to the SEM's bias foils was typically 100~Volts, greater than the typical 20-30~eV kinetic energy of secondary electrons emitted from a foil surface.\cite{bruining}  To understand the effect of the bias voltage on the signal collection efficiency, we accumulated data at several fixed beam intensities varying from $4\times10^{11}$~protons/pulse (ppp) to $4\times10^{12}$~ppp.  During each period of fixed beam intensity, the voltage applied to the bias foils was varied and the total charge collected from all the signal strips measured.  The results are plotted in Figure~\ref{fig:biasvolts}.  The signal collected from the SEM is normalized to 100\% efficiency by dividing by the signal collected at $4\times10^{12}$~ppp and 100~Volts.  Both the horizontal and vertical signal foils agree within 1.5\%.  As can be seen, applying a voltage increases the efficiency, as has been noted by others \cite{isabelle,blankenburg}.  Also, our data suggest that the required applied voltage to achieve 100\% efficiency increases as the beam intensity increases.  The magnitude of this effect may reflect surface impurities on the foils or be caused by the relatively poor vacuum in this chamber, which was several 10$^{-7}$~Torr during the test.

\section{Measured SEM Centroid Resolution}
\label{sec:SEMprec}
  
The centroid resolution of the SEM depends upon the width of the beam, the intensity of the beam, and upon the electronics readout noise on the SEM channels. The beam width in vertical direction was nearly constant at $\sigma_x=3.4$~mm, while the horizontal beam size varied between $\sigma_y=7.4$~mm and $\sigma_y=8.2$~mm, during the test period. This gives us 3 different beam widths for which we try to find the SEM centroid resolution. 

\begin{figure}
  \begin{center}
  \includegraphics[width=0.48\textwidth]{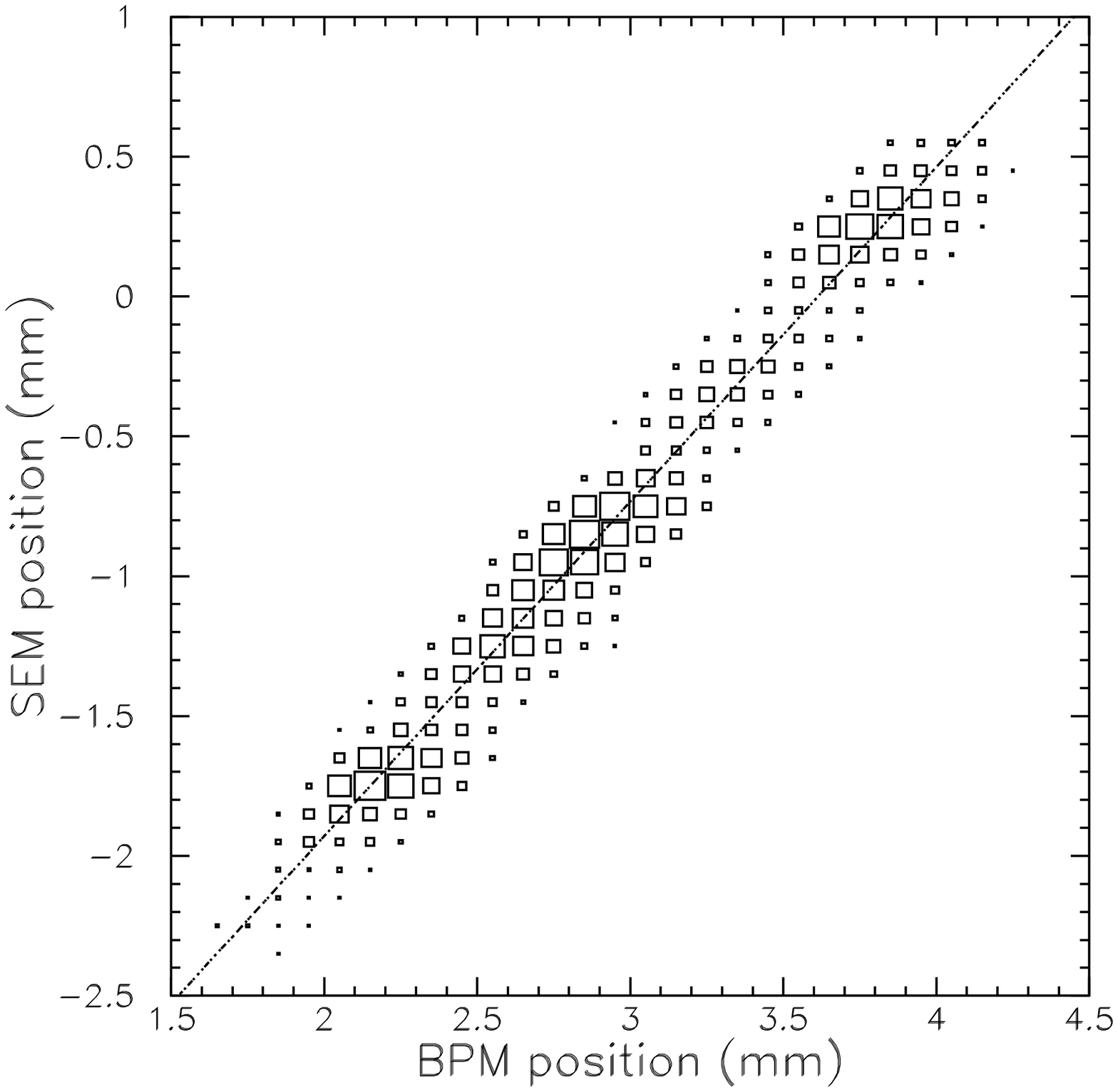}
  \includegraphics[width=0.48\textwidth]{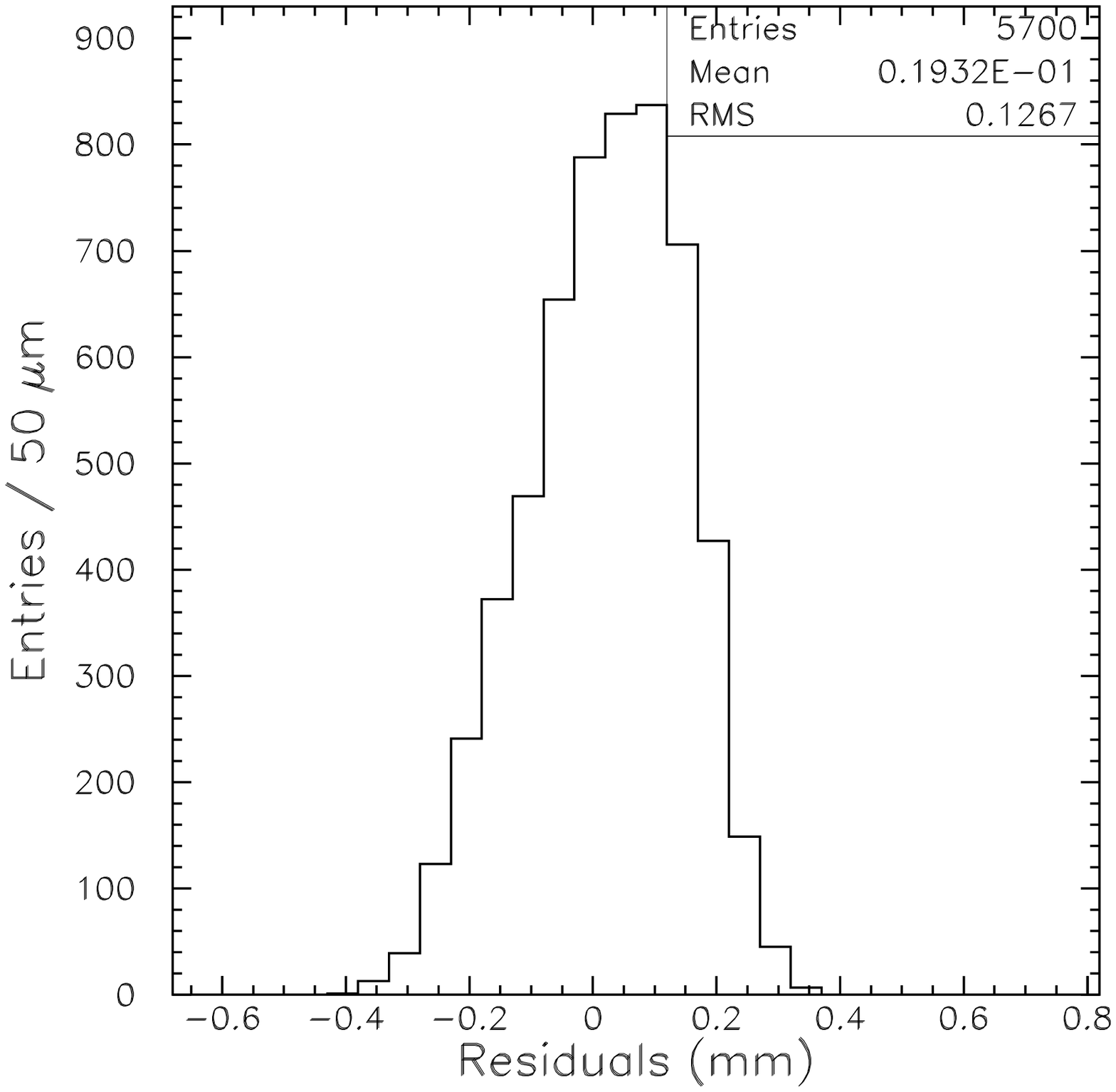}
  \caption{(left) The correlation between positions measured by the vertical BPM and those measured by the SEM.  (right) The residuals from the best fit line.   From the residuals we can infer that the sum in quadrature of BPM and SEM resolutions is 127$\mu m$.
  \label{fig:beammove}}
  \end{center}
\end{figure}

To find the beam centroid resolution in the vertical plane we correlate its reported beam position with that of the BPM labelled VP871. Figure~\ref{fig:beammove} shows the BPM data ploted versus SEM data over a range of spills in which the beam was observed to move substantially across the chambers. The residuals of the data from the best fit line are also shown in a figure, and show an RMS of 127~$\mu$m.  The residuals should be a measure of the two devices' resolutions, added in quadrature:

\begin{equation}
\sqrt{\sigma_{SEM}^2 + (\alpha \sigma_{BPM})^2} = 127~\mu m.
\label{eq:scatter}
\end{equation}

where $\sigma_{SEM}$ and $\sigma_{BPM}$ are the intrinsic resolutions of the detectors, and $\alpha=1.1$ is a scale factor between the SEM and BPM which is taken from the slope of the line in Figure~\ref{fig:beammove}. This scale factor can result from either optics of the transport line or miscalibrations between the SEM and BPM.  

\begin{figure}
  \begin{center}
  \includegraphics[width=0.48\textwidth]{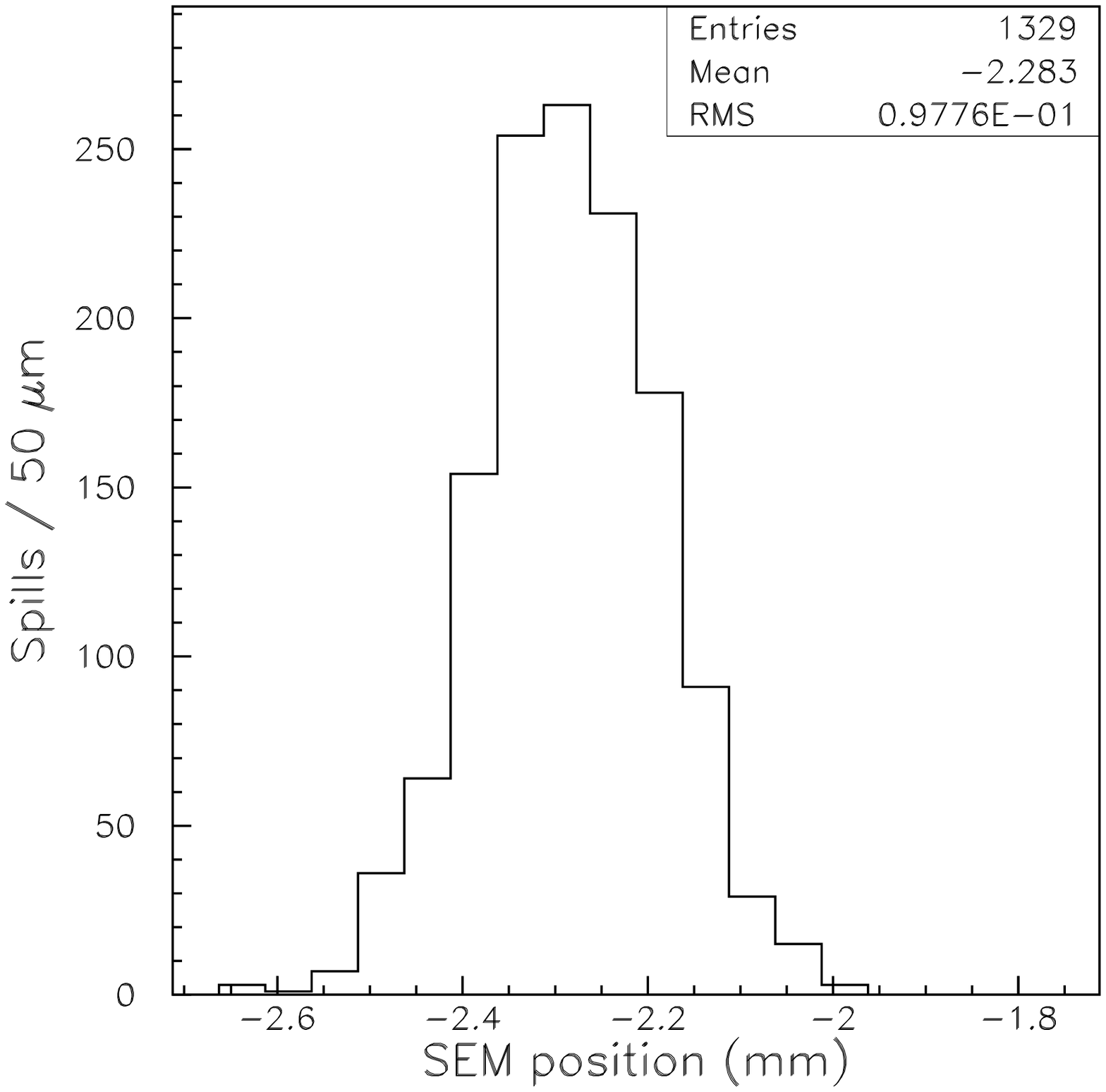}
  \includegraphics[width=0.48\textwidth]{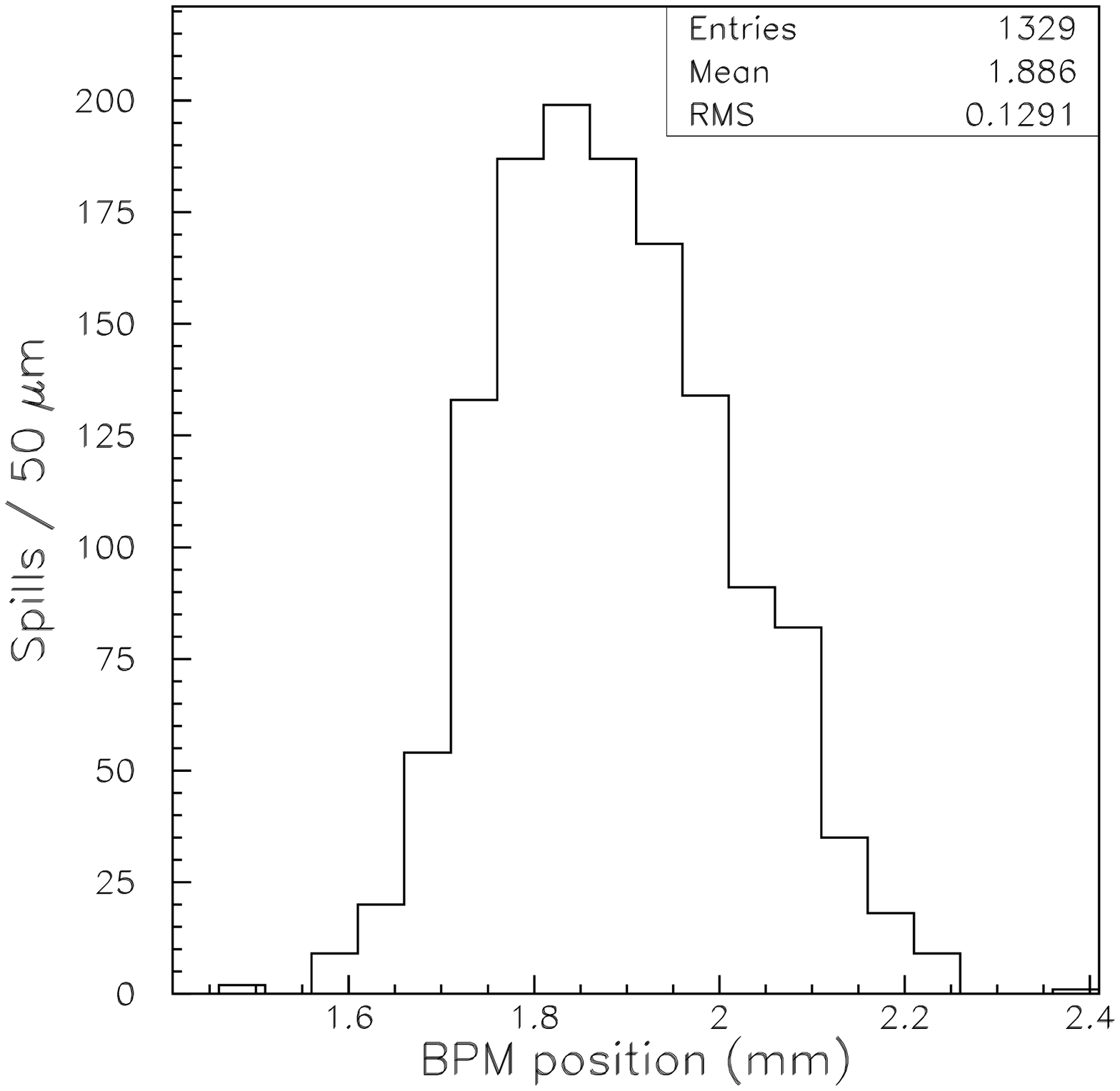}
  \caption{Comparison of the reported beam position from the SEM and BPM during a time period of 1329 spills when beam position is nearly constant. 
  \label{fig:stillbeam}}
  \end{center}
\end{figure}

In order to separate the individual BPM and SEM resolutions, we select a different time span of spills in which the beam position at the detectors was observed to be consistent from spill to spill (to within $<$0.5mm).  Figure~\ref{fig:stillbeam} shows a histogram of the beam positions, as measured in each of the detectors, for one such interval of 1329 spills in which the beam motion was relatively small. The RMS of the beam centroid positions spill-to-spill is a measure of both the beam wandering and the device resolution, {\it ie.}
\begin{eqnarray}
\label{eq:wander}
RMS_{SEM} = \sqrt{\sigma_{SEM}^2 + \sigma_{wander}^2} = 97.8 \mu m\\
RMS_{BPM} = \sqrt{(\alpha \sigma_{BPM})^2 + \sigma_{wander}^2} = 129.1 \mu m \nonumber
\end{eqnarray}
where $\sigma_{wander}$ is a measure of the RMS variation of the beam motion spill-to-spill due to variations in the beamline performance, $\sigma_{BPM}$ is the intrinsic resolutions of the BPM, $\sigma_{SEM}$ is the intrinsic resolutions of the SEM, and $\alpha$ is the scale factor to account for differences between the SEM and BPM calibrations.  The quantity $\sigma_{wander}$ is not {\it a priori} known and varies from time interval to time interval but it should be the same for both devices because of their proximity.  As may be seen in Figure~\ref{fig:stillbeam}, the BPM spread is larger than the SEM spread, suggesting $\sigma_{SEM}<\sigma_{BPM}$.\footnote{In fact, we selected over 20 such time spans of relatively stable beam and found that the spill-to-spill variation reported by the BPM was greater than that reported by the SEM.} 

Combining the expressions in Equation~\ref{eq:scatter} and Equation~\ref{eq:wander}, and using $\alpha=1.1$ for the scale factor between BPM and SEM positions, we may obtain:
\begin{center}
$\sigma_{SEM}=67 \pm 5 \mu$m \\
$\sigma_{BPM}=98 \pm 5 \mu$m
\end{center}
We have repeated this analysis of the vertical beam data for several time intervals.  The resolutions are observed to vary by an RMS of 5~$\mu$m, with the variation possibly due to beam-related effects or variation in resolution across the aperture of the SEM or BPM.

In the horizontal view the presence of the focusing quadrupoles 870A and 870B, as well as the trim magnet HT870, complicates direct comparison of the beam position reported by the prototype SEM and the nearby horizontal BPM labelled HP870.  Furthermore, the beam width varied between 7.4 mm and 8.2 mm, and was correlated with two different beam intensities. The expected SEM centroid resolution is quite different for those two beam widths, so we split data into two sets. We looked at 32 intervals with 1000 spills each. We assumed that the resolution for the horizontal BPM HP870 is the same as for the identically-constructed vertical BPM VP871. In each time interval we could find beam wandering ($\sigma_{wander}$) from the BPM measurements and then plug that into SEM data and find $\sigma_{SEM}$. As a result we find:
\begin{center}
$\sigma_{SEM}(\sigma_x=7.4mm)=151 \pm 5 \mu$m\\
$\sigma_{SEM}(\sigma_x=8.2mm)=171 \pm 5 \mu$m
\end{center}  

\section{Expected SEM Resolution}
\label{sec:measacc}
The previous section measured the centroid resolution at three different beam widths.  A less reliable estimate was also made of the beam width resolution (see below).  The beam centroid and width resolutions of a SEM grid are affected by several instrumental factors\cite{PlumBIW04}:
\begin{itemize}
\item The signal noise on each individual SEM strip ($\delta y_i$).
\item The non-uniformity in strip spacing ($\delta x_i$) 
\item The number of SEM strips per beam $\sigma$
\end{itemize}
The first effect listed above smears the pulse height $y_i$ observed on an individual strip $i$ by an amount $\delta y_i = \sqrt{\delta^2 + (\varepsilon y_i)^2}$, where $\delta$ is a presumed constant pulse height resolution and $\varepsilon$ describes a signal size-dependent resolution.  The second effect, which arises due to the fact that the foil strips are not perfectly positioned on the grid, causes a smearing of the actual position $x_i$ of the strips. This form of strip placement error we accounted for by assuming an additional uncertainty of the strip pulse heights of $\delta y_i=f'(x)\cdot \delta x_i$ where $f'(x)$ is the derivative of the gaussian fitting function that we use to describe the beam profile and $\delta x_i$ is the uncertainty in the strip positions. 

We fit our beam test data to a model of expected beam centroid resolution which incorporates the above effects.  We simulated beam spills in a 44-channel detector whose signals were smeared to account for electronics noise and foil strip misalignments.  Such simulated data was generated for different beam widths, and for different assumptions on the electronics noise and strip misalignments.  By comparing to our data, we could in effect measure the electronics noise.  

\begin{figure}[t]
  \begin{center}
  \includegraphics[width=.85\textwidth]{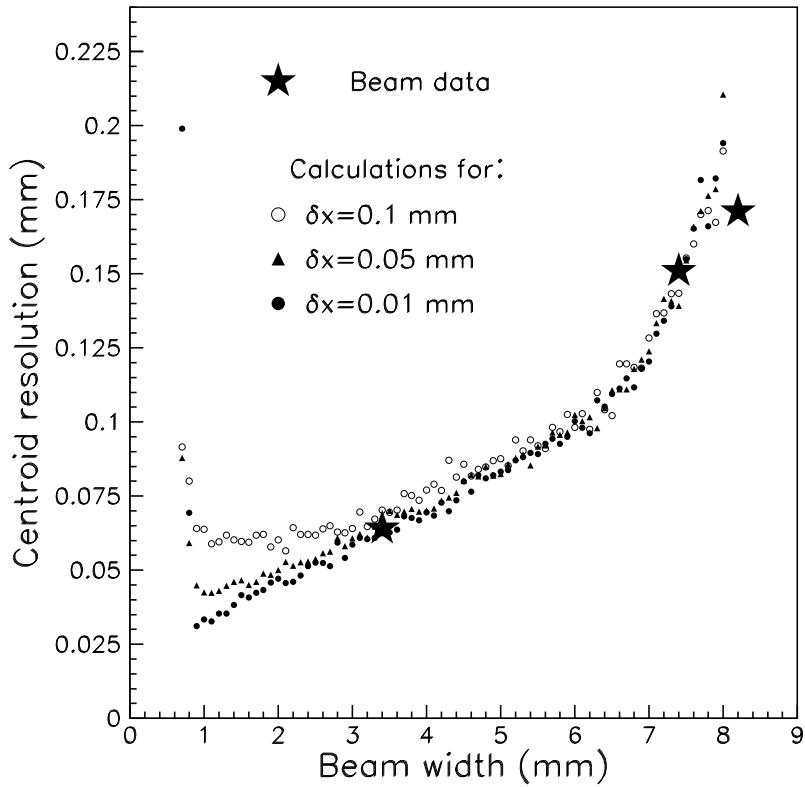}
  \caption{Comparison of the measured resolution of the centroid position from this beam test to a calculation based on simulated data for different beam widths. The quantity $\delta x$ is the amount by which the foil strips are misplaced on the grid.  Constant intensity is assumed in the calculation.
  \label{fig:errmeanvssgm}}
  \end{center}
\end{figure}

\begin{figure}[t]
  \begin{center}
  \includegraphics[width=.85\textwidth]{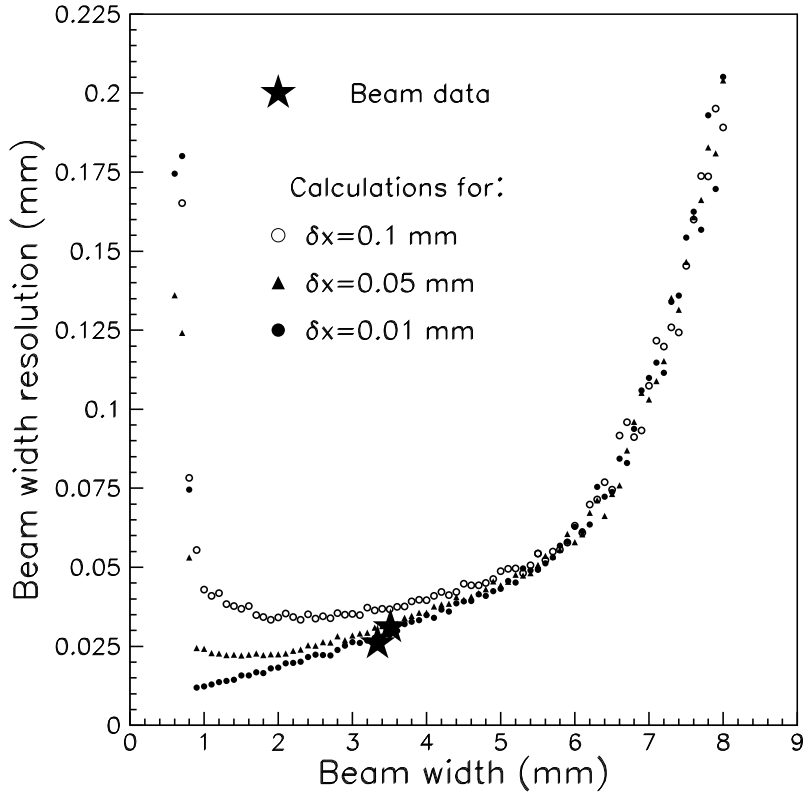}
  \caption{Comparison of the measured resolution of the beam width from this beam test to a calculation based on simulated data for different beam widths. The quantity $\delta x$ is the amount by which the foil strips are misplaced on the grid.  Constant intensity is assumed in the calculation.}
\label{fig:errwid}
  \end{center}
\end{figure}

Our test beam data is overlaid with the corresponding calculations in Figure~\ref{fig:errmeanvssgm} and \ref{fig:errwid}.  As can be seen, we anticipate smaller centroid and width resolutions for narrower beams, until the beam width $\sigma_{beam}=1.0$~mm, which corresponds to the strip-to-strip spacing for our SEM.  As the beam becomes narrower, the centroid and width resolutions are more sensitive to the placement accuracy $\delta x_i$ of the strips in the SEM grid.  Our calculation is performed for three different accuracies, $\delta x_i=10, 50,$ and $100~\mu$m.  We expect $\delta x_i=50~\mu$m for the prototype and $\delta x_i=10~\mu$m will be achieved for the final NuMI SEM's. Figure~\ref{fig:errmeanvssgm} indicates a centroid resolution of order 20-30~$\mu$m may be anticipated in the NuMI beam line, where the beam width $\sigma_{beam}$ is about 1.0~mm.  

We have also tried to understand the expected resolution of the beam width measured by the SEM grid.  Figure~\ref{fig:errwid} shows the expected resolution of the beam width as a function of the beam width.  
Overlaid on the plot in Figure~\ref{fig:errwid} are two points derived from the MiniBooNE test beam run which are the variation of the width in the vertical plane as observed over two ranges of beam spills.  A similar procedure was not possible in the horizontal, as the beam width was observed to vary dramatically at high beam intensity to to emittance variation from the 8 GeV Booster accelerator, an effect confirmed by other instrumentation in the transport line.  


\section{Conclusion}
From the prototype data we observe that the beam centroid resolution of the 1~mm pitch SEM prototype is around $64~\pm~5~\mu$m for the beam with $\sigma=3.5$~mm.  From the extrapolation in Figure~\ref{fig:errmeanvssgm} to beam widths relevant for NuMI, we anticipate a centroid resolution of  20-25~$\mu$m for a 1~mm beam. Although the beam intensity will be a factor of 5 larger in the NuMI beam, which might be expected to improve the SEM resolution due to increased signal size, this signal increase will be compensated by a signal decrease arising from the narrower foil strip size in the NuMI SEM's.  Thus, the extrapolation shown in Figures~\ref{fig:errmeanvssgm} and \ref{fig:errwid} should be approximately correct.  

The two SEM's just upstream of the NuMI target will have finer pitch than the prototype SEM (0.5~mm compared to 1.0~mm), and also wider strips (0.25mm as compared to the 0.15mm of the transport line SEM's).  One therefore expects that (a) the mechanical assembly details of the 0.5mm SEM's shall not be as critical, since the beam size will be larger than the strip spacing, and (b) the pulse height smearing will not as greatly affect their resolutions because the wider strips will yield a 1.7 times greater signal.  The results of the present study suggest that these 0.5~mm SEM's should behave in a 1.0~mm beam much like the 1.0~mm prototype performance for a 2.0~mm beam.  That is, an approximate scaling relation should exist between the strip pitch and the beam width.

\section{Acknowledgements}
We thank Gianfranco Ferioli of CERN for extensive advice and consultation on segmented SEM's and for sharing his designs,
upon which the present prototype is based.
The beam test described in this memo was performed thanks to the efforts of members of the FNAL Particle Physics and Accelerator Divisions as well as the University of Texas Department of Physics Mechanical Support Shops.  This work was supported by the U.S. Department of Energy under contracts DE-FG03-93ER40757 and DE-AC02-76CH3000, and by the Fondren Family Foundation.




\begin{thebibliography}{00}
\bibitem{caveat} The SEM pulse height is nominally a charge whose magnitude should be of order a percent of primary beam charge traversing a foil strip.  These pulse heights were read out using charge-integrating amplifiers whose gain and charge-to-voltage conversion were not well known for our test.  The lack of knowledge of this conversion factor is not important for our final results.
\end{thebibliography}
\end{document}